\documentclass[review]{elsarticle}
\usepackage{bm}%
\usepackage{graphicx}
\usepackage{dcolumn}
\usepackage{amssymb}

\journal {Fluid Phase Equilibria}
\begin{document}
\begin{frontmatter}
\title{Molecular Dynamics Simulation of Liquid-Vapor Phase Diagrams of Metals Modeled Using Modified Empirical Pair Potentials}
\author{A. Sai Venkata Ramana}
\address{Theoretical Physics Division, Bhabha Atomic Research Centre, Mumbai-400085, India}
\date{\today}
\begin{abstract}
We propose a modified form of pair potential for metals. The parameters of the potential 
 are obtained by fitting the cold curve of the potential to that obtained from the ab-initio 
calculations. Parameters have been obtained for Aluminum, Copper, Sodium and Potassium.
To test the accuracy of the potentials, we performed particle-transfer molecular dynamics
 simulations and obtained the liquid-vapor coexistence curves of the above metals.
 We found that, in the cases of Sodium and Potassium, the present results improve significantly over those 
 obtained from Morse potential (J.K. Singh et. al., Fluid Phase Equilibria 248(2006)). 
 In the cases of Aluminum and Copper, the present results are closer to those obtained from the Morse potential.
 We also obtained isobars of Aluminum and Copper at 0.3GPa from NPT ensemble simulations.
We observed that the isobars obtained using the Morse potential and the modified potentials are in close agreement
in both the cases. The obtained isobar of Copper is in reasonable agreement with the  
 experimental isobar while that of Aluminum is slightly deviating from the experimental isobar.
\end{abstract}
\end{frontmatter}

\section{Introduction}
Equation of State(EOS) of materials under wide ranges of temperatures, pressures and densities is an essential input to 
hydrodynamics simulations which find applications in astrophysics, high energy density physics etc.
In the compression regime, ample of experimental data is available for most of the materials of interest and
 a number of accurate theoretical models of EOS which have been tested against experimental data are available\cite
{qeos1,qeos2,qeos3,qeos4}. But in the expansion regime, particularly in the case of metals (except alkali metals),
 reliable experimental data is not available because of the drastic temperature and pressure conditions involved.
 On the other hand, ab-initio simulations, even though 
being accurate, require enormous amount of computation time and it is impractical to
 generate EOS libraries using these methods.
 Thus theoretical methods based on classical statistical mechanics such as thermodynamic perturbation theory(TPT),
 integral equation theory(IET) etc.\cite{Hansen}, still form 
important tools to obtain thermodynamic properties of materials in fluid phase.
 Accuracy of these methods depend on two factors; first is the approximation(s) used in the theory, and second the 
 inter-particle potential which characterizes the material. Given the inter-particle potential, the presently available methods to determine 
 thermodynamic properties of fluids are  seen to be reasonably accurate\cite{TPT1,TPT2,TPT3,TPT4}.
  Thus the agreement of EOS and phase diagrams in fluid phase with experimental data or ab-initio data mainly depends on
  the accuracy of the inter-particle potential.

Presently various sophisticated potentials based on embedded atom model (EAM)
\cite{Daw1,Daw2} are available in the literature.
 Particularly in the case of metals various potential forms are available\cite{Zhou, Sutton, Cleri,Gelb,Alek0, Bour}
which have been seen to perform well in predicting various physical properties in solid phase, melting, boiling, 
 surface tension etc.
However, the theoretical methods like TPT, IET etc. require only inter-particle pair potentials.
Two model pair potentials are widely used to model the inter-particle interactions in metals. They are the generalized Lennard-Jones 
(GLJ)\cite{GLJ} and the Morse potentials\cite{Morse}.
Recently, Sun Jiuxun\cite{Jiu} obtained the parameters of a modified GLJ potential for various 
materials and showed that the zero Kelvin pressure isotherm obtained from this potential is quite accurate. 
However, we observed that the zero Kelvin energy isotherm obtained from the potential is erroneous. Also we noticed that the modified GLJ potential 
with parameters as obtained by Sun cannot be used in simulations and in theoretical methods for most of the materials\cite{sai}.
 The parameters of Morse potential for various cubic metals have been obtained by Lincoln et. al. \cite{Lin}
 using the Lattice parameter, bulk modulus and cohesive energy data.
 J.K. Singh et. al. \cite{Singh1} obtained the liquid-vapor phase diagrams of various metals modeled using the Morse 
potential as parametrized by Lincoln et. al. through grand-canonical Monte-Carlo simulations.
 Their results show that there is enormous deviation between the simulation and experimental values in the case of alkali metals. 
The deviations would have arisen because of obtaining the potential parameters using only few 
physical quantities like cohesive energy, bulk modulus and lattice parameter which are just equilibrium properties.

Our hunch is that, a potential whose cold curve (zero Kelvin isotherm) matches with the experimental cold curve
 or the one obtained from accurate calculations based on density functional theory(DFT) etc., 
would predict thermodynamic properties accurately over a wide range of thermodynamic conditions. 
On the other hand, accurate and fast DFT calculations of the cold curve are possible with the present day available codes. 
So in the present work, we use the cold curve obtained from ab-initio calculations to obtain parameters of the potential for various metals.
 We fit the cold curve obtained from a given form of potential to that obtained from ab-initio data of the required
 metal. This, in-addition to the potential parameters, gives a clear idea of the 
accuracy of the cold curve corresponding to a potential away from the equilibrium.

 When the above mentioned procedure has been done using the Morse and GLJ potentials, we observed that the cold 
curves obtained from them do not accurately fit the ab-initio data away from the equilibrium, particularly in the case of alkali metals.
Hence we propose a modified form of empirical pair potential for metals; the cold curve of which is expected to fit the ab-initio data accurately.
 Parameters of the modified potential have been obtained for Aluminum, Copper, Sodium and Potassium.
To test the accuracy of the potentials obtained, we obtained the liquid vapor coexistence points for various 
temperatures for the metals considered using particle transfer molecular dynamics(PTMD)\cite{Lu1,Lu2,Lu3} 
simulations. The PTMD is a method to simulate the liquid vapor coexistence conditions which has been developed taking
inspiration from Gibbs Ensemble Monte Carlo method originated by Panagiotopoulos\cite{Pan0}.
Also we have done NPT ensemble simulations to obtain isobars of Aluminum and Copper at 0.3GPa to test the accuracy
of the present potentials at lower temperatures.

 In  Section $2$, we describe our method of obtaining parameters of the potentials using ab-initio cold curve.
 In section $3.1$, the PTMD method
  we adopted is explained in brief and liquid-vapor phase diagrams(LVPDs) we obtained for Aluminum, Copper, Sodium and Potassium are 
compared with available data.
In section $3.2$, NPT ensemble simulations to obtain isobars of Aluminum and Copper are discussed.
The results are analyzed in section $3.3$ and the paper is concluded in Section $4$.

\section{Pair Potential Models for Metals}
The GLJ potential is given by
\begin{equation}
u_{LJ}(r) = \frac{\epsilon}{m_1 -n_1}\left[ n_1\left( \frac{r_0}{r}\right)^{m_1}   - m_1\left( \frac{r_0}{r}\right)^{n_1}      \right] \label{glj}
\end{equation}
\noindent
Jiuxun\cite{Jiu} proposed a relation between $m_1$ and $n_1$ so that only one is independent.
The relations are $m_1 = 6n-3$ and $n_1 = 3n-3$. Taking his modification into account, the potential becomes,
\begin{equation}
u_{LJ}(r) = \frac{\epsilon}{3n}\left[ (3n-3)\left( \frac{r_0}{r}\right)^{6n-3}   - (6n-3)\left( \frac{r_0}{r}\right)^{3n-3}      \right] \label{mglj}
\end{equation}
The Morse potential is
\begin{equation}
u_M(r) = \epsilon(e^{-2\alpha(r - r_0)} - 2e^{-\alpha(r - r_0)})   \label{morse}
\end{equation}
\noindent
We propose a new form of pair potential given by
\begin{equation}
u_s(r) = \epsilon\left[ e^{-2\alpha(r/r_0 - 1)}\left( \frac{r_0}{r}\right)^{2\beta}  -  2e^{-\alpha(r/r_0 - 1)}\left( \frac{r_0}{r}\right)^{\beta}            \right]    \label{sai}
\end{equation}
Attractive part of the above potential is chosen inspired by the screening of ions by electrons in metals. Repulsive part is chosen ad-hoc as
 per mathematical convenience. It can be seen that, by putting $\beta=0$ in above potential we recover the Morse potential and by 
putting $\alpha=0$,it becomes a Lennard-Jones type potential. Since the electron-ion screening term is represented by an exponential term
multiplied by the coulomb term (in the semi-classical picture), we expect that a similar form with the exponents of the
exponential term and the Coulomb term being adjusted, would represent the interactions in a better way.
 
 In the case of GLJ potential, $r_0,n$ and $\epsilon$ are the parameters. 
For the Morse potential, $\epsilon, \alpha$ and $r_0$ are parameters and for the modified potential, $\epsilon, \alpha, \beta, r_0$ 
are the parameters. For all the potentials, the potential is minimum at $r_0$ and the well depth is $\epsilon$.

\subsection{Obtaining the Potential Parameters}
We used Energy per particle Vs volume per particle data obtained from ab-initio calculations to obtain the parameters of the potential for various metals. Ab-initio calculations were done using VASP\cite{Vasp1,Vasp2,Vasp3,Vasp4} software.
 PAW potentials and Monkhorst-Pack grid in reciprocal space have been used. The k-
point grid has been adjusted inspecting the convergence of the total energy in each case. 
Energy cutoff used is $400 eV$ and the energy convergence criterion is set to be $10^{-6} eV$.
Also, the energy of free atom has been calculated by
taking a large volume per atom and has been subtracted from the total energy per
particle so that cohesive energy can be calculated directly from the curve.
The cold curve for each model potential is written as follows.
\begin{equation}
U_x = \sum_{i=1}^{m} \frac{\delta_i \epsilon}{2} u_x(a_i)  \label{ulat}
\end{equation} 
\noindent
where $x$ can be $LJ,M$ or $s$. $U_x$ is the energy per particle. $a_i$ is the distance of the $i^{th}$ neighbor from a particle situated at origin.
 $\delta_i$ is the number of $i^{th}$ neighbors. In the present work we have accounted interaction up to $(m=)10^{th}$ neighbor shell. 
 Now $a_i$ is related to volume per atom as follows: 

In the case of a fcc solid, $a_i = \sqrt i a_1$ and $a_1 = a/\gamma$. Where $a$ is the lattice parameter and $\gamma$ is the structural 
constant. For fcc solids, volume per atom $V=a^3/4$ and $\gamma$ is equal to $\sqrt 2$. For simplicity, we write $r_0$ as 
$(4V_0)^{1/3}/\gamma$ without loss of generality. Thus using this information, Eq.(\ref{ulat}) can be written in terms of 
volume per atom $V$. The equation then becomes

\begin{equation}
U_x = \sum_{i=1}^{m} \frac{\delta_i \epsilon}{2} u_x(V/V_0)  \label{ulat2}
\end{equation} 
\noindent
For example, using $u_s(r)$ in $U$,
\begin{equation}
U_s = \sum_{i=1}^{m} \frac{\delta_i \epsilon}{2}\left[ \frac{e^{-2\alpha(\sqrt i(V/V_0)^{1/3} - 1)}}{i^{\beta}}\left( \frac{V_0}{V}\right)^{2\beta/3}  -  2\frac{e^{-\alpha(\sqrt i(V/V_0)^{1/3} - 1)}}{i^{\beta/2}}\left( \frac{V_0}{V}\right)^{\beta/3}           \right]  \label{uvfcc}
\end{equation} 

In the case of a bcc solid, $\gamma= 2/\sqrt 3$ and $V= a^3/2$. However, the $a_i$ do not hold a general relation with $a$ as for fcc solids 
and have to be carefully calculated. In this case $r_0$ has been chosen as $(2V_0)^{1/3}$. Writing $a_i = d_i a$ where $d_i$ has to be
calculated using the crystal structure, equation corresponding to $u_s(r)$ is
   
\begin{equation}
U_s = \sum_{i=1}^{m} \frac{\delta_i \epsilon}{2}\left[ \frac{e^{-2\alpha(d_i(V/V_0)^{1/3} - 1)}}{d_i^{2\beta}}\left( \frac{V_0}{V}\right)^{2\beta/3}  -  2\frac{e^{-\alpha(d_i(V/V_0)^{1/3} - 1)}}{d_i^{\beta}}\left( \frac{V_0}{V}\right)^{\beta/3}           \right]  \label{uvbcc}
\end{equation} 
\noindent

\begin{figure}
  {\includegraphics[scale=0.6,angle=-90]{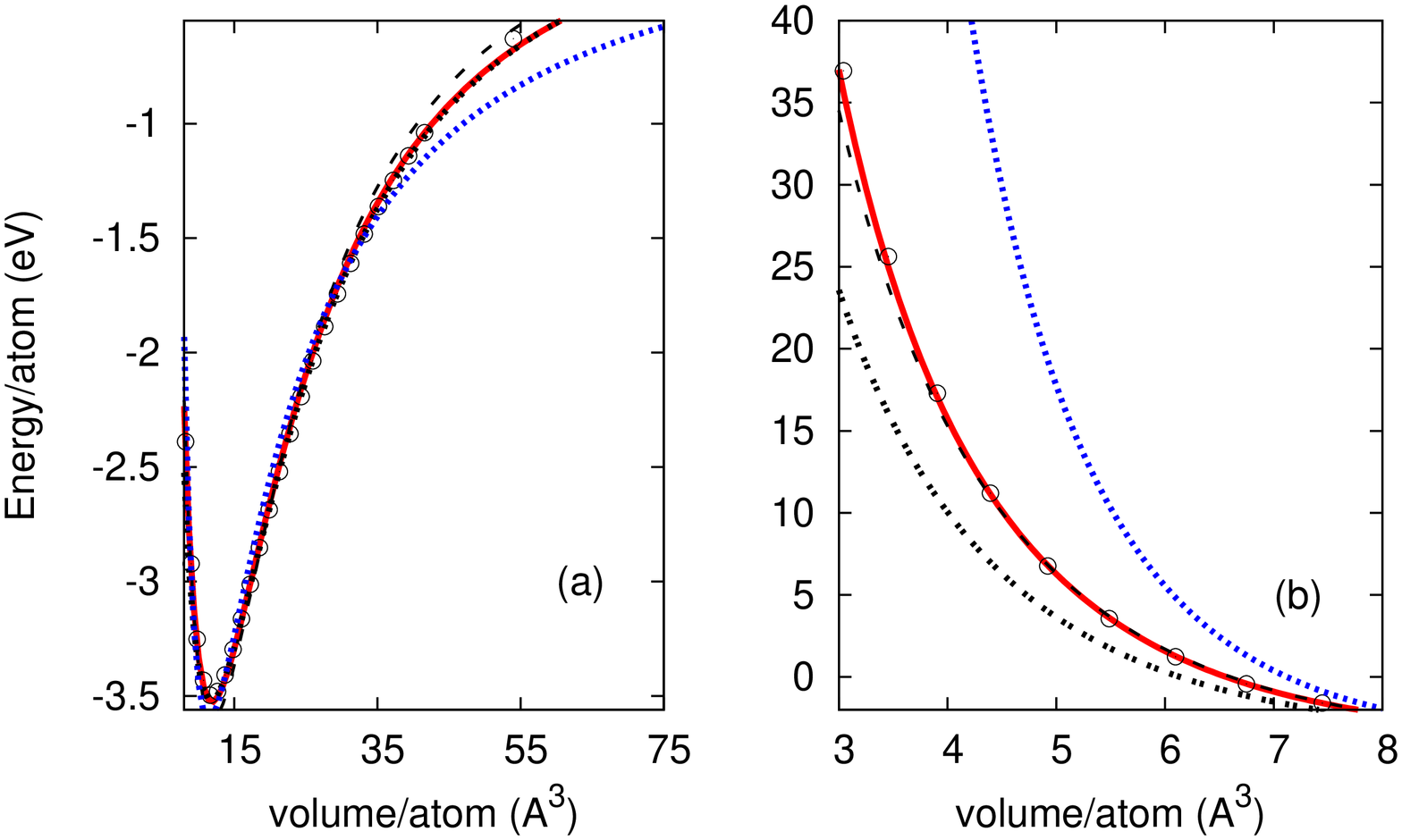}}
 \caption{\label{2}}
 \end{figure}

\begin{figure}
 {\includegraphics[scale=0.6,angle=270]{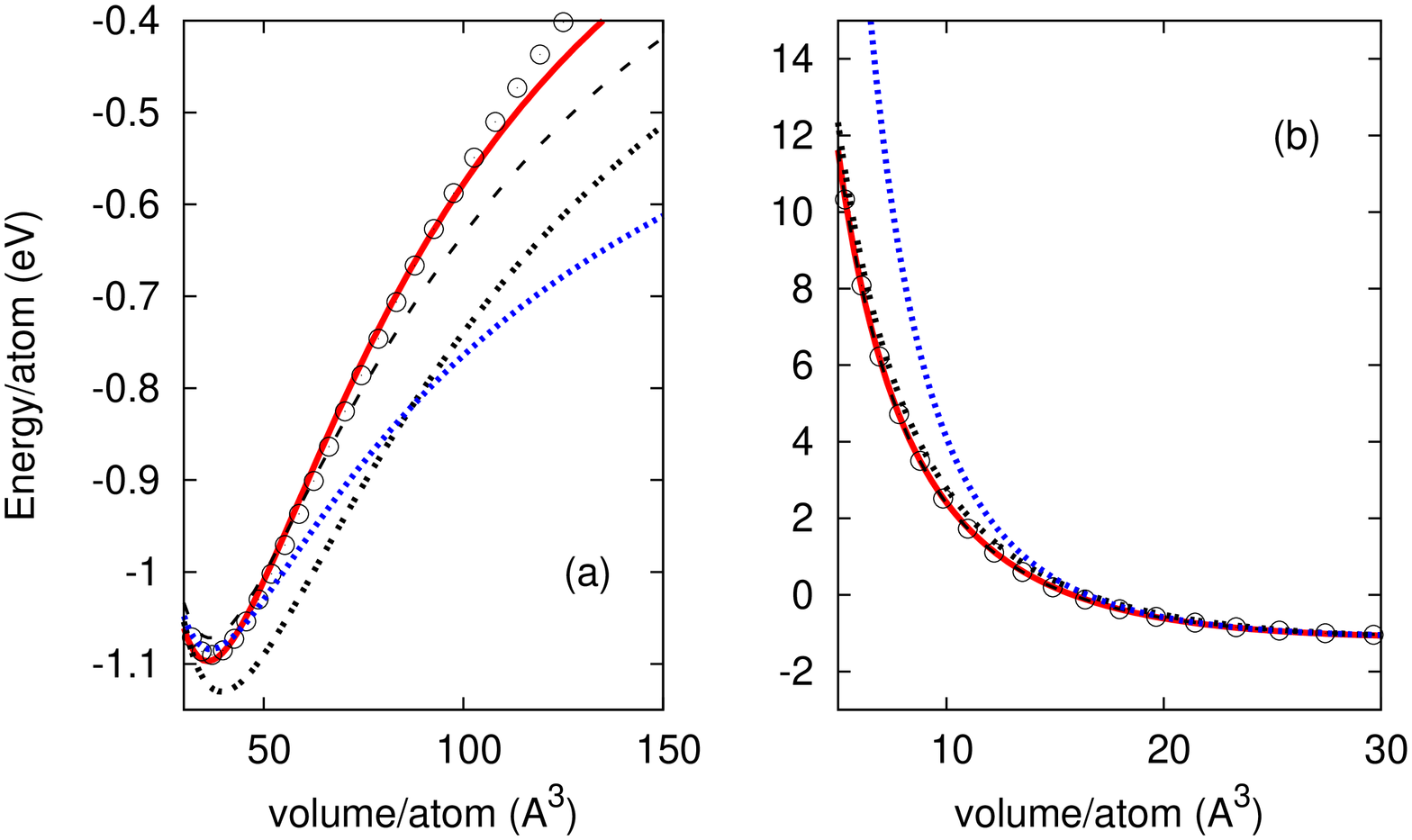}}%
 \caption{\label{3} }
\end{figure}

\begin{figure}
  {\includegraphics[scale=0.6,angle=270]{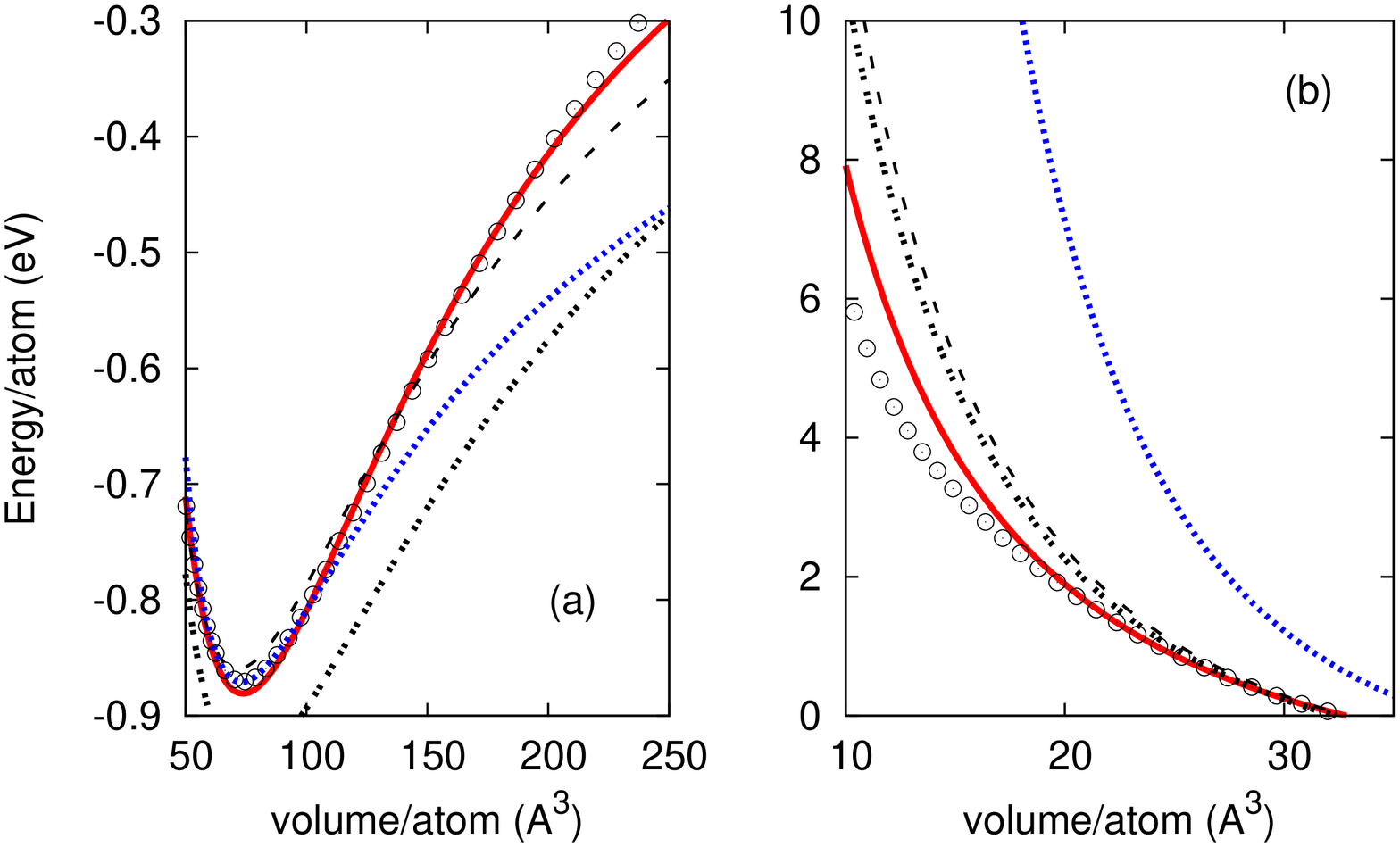}}%
 \caption{\label{4} }
 \end{figure}

The parameters of each potential model are obtained by fitting the cold curve Eq.(\ref{ulat2}) to the ab-initio cold curve. 
The fitted zero Kelvin isotherms obtained from various potentials models for Aluminum, Copper, Sodium and
 Potassium are shown in Figs.(\ref{1}-\ref{4}). From the figures it can be seen that the cold curve obtained from GLJ potential is quite off from 
the ab-initio data away from the equilibrium in all the cases, in both compression and expansion
phases. The cold curve obtained from Morse potential with parameters of Lincoln et. al.($U_{ML}(V)$), in the case 
of Aluminum has a slight deviation from ab-initio data near the equilibrium and away from equilibrium also, as the cohesive energy predicted by
ab-initio calculations and that used by Lincoln et. al. differ slightly. 
On the other hand, $U_{M}(V)$ and $U_s(V)$ 
fit the ab-initio data with same accuracy except for a slight deviation of $U_M(V)$
 from the ab-initio data away from equilibrium in the expansion phase.
In the case of Copper, $U_{ML}(V)$, $U_{M}(V)$ and $U_s(V)$ are equally accurate in the expansion phase. But in the 
compression phase, $U_{ML}(V)$ has a significant deviation from the ab-initio data.

In the cases sodium and potassium,  $U_{ML}(V)$ is deviating significantly from ab-initio 
data. This could be because of improved accuracy of the present
ab-initio calculations in predicting the cohesive energies.
 Also, the $U_M(V)$ is deviating from the ab-initio data in the expansion
 region far from the equilibrium for both sodium and potassium.
 Whereas in the compression region, $U_M(V)$ is matching reasonably well with 
ab-initio data in the case of Sodium and is deviating from ab-initio data in the case of  Potassium.
It can be seen that $U_s(V)$ matched reasonably well with ab-initio cold curve in both the cases.

\begin{figure}
 {\includegraphics[scale=0.6,angle=270]{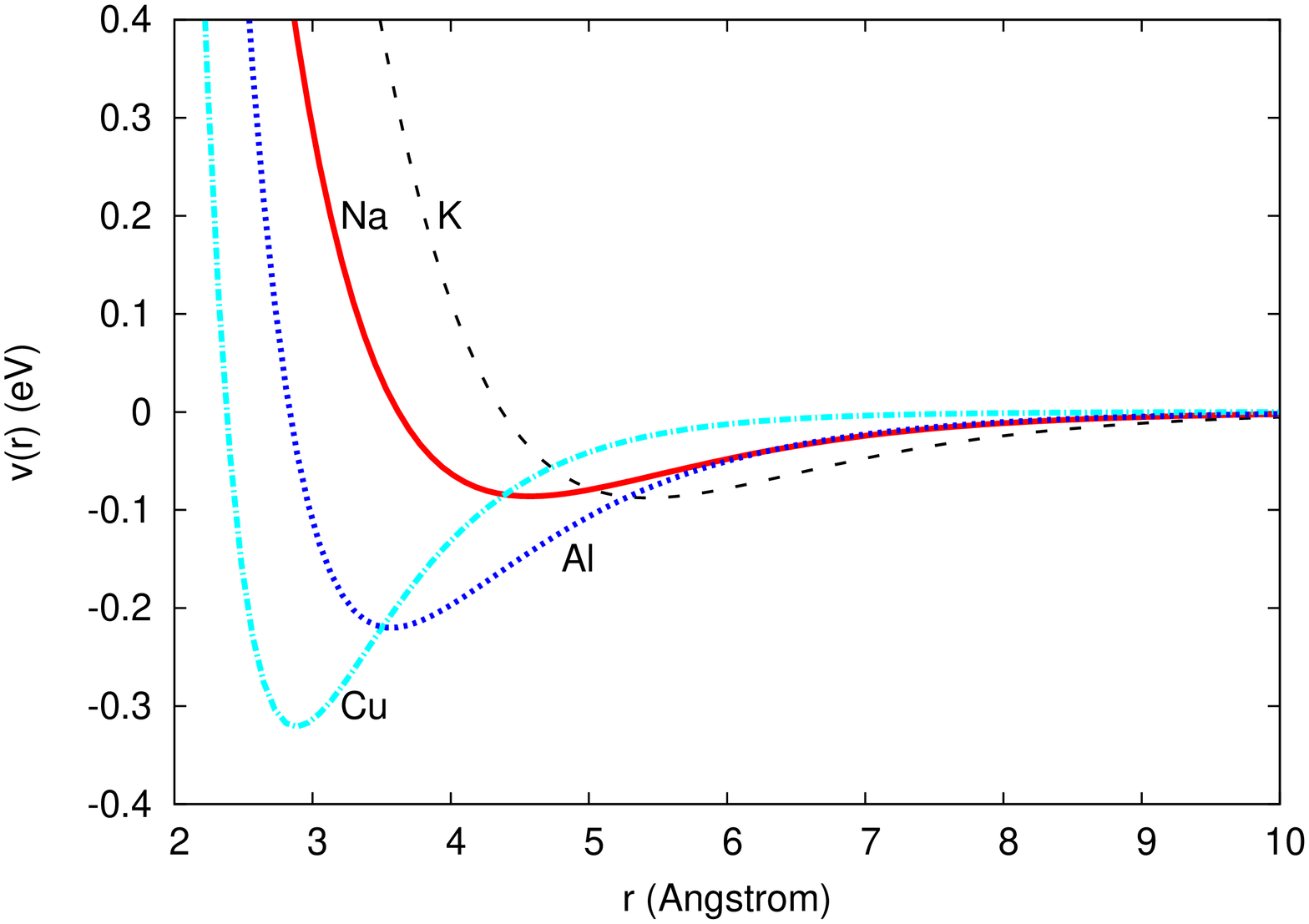}}%
 \caption{\label{5} }
\end{figure}

Above discussion shows that $U_s(V)$ obtained from $u_s(r)$ fitted the ab-initio cold curve with better accuracy than
 those of other potentials in all the cases shown.
 Thus we expect that $u_s(r)$ would give an improved description of the metals in the 
liquid-vapor coexistence region. $u_s(r)$ for Aluminum, Copper, Sodium and potassium is shown in Fig.(\ref{5}).
Parameters for the potential $u_s(r)$ obtained using the procedure described above are listed in Table.(\ref{11}).
For Sodium and Potassium, the parameter $\beta$ becomes negative which makes the potential turn down and go
to zero close to origin which is un-physical.
 Thus it has to be cutoff at an appropriate point close to the origin. We found that $0.25r_0$
 and $0.4r_0$ can be the cut-off points for Sodium and Potassium respectively. It is assumed that for
 distances smaller than these, the potential is constant and equal to that at the cutoff point.
 Clearly, this would not affect the results as the probability of finding
 a particle inside the repulsive core is negligible.
In order to test the accuracy of modified potential $u_s(r)$ in the fluid phase,
 we have obtained liquid-vapor phase diagrams
(LVPDs) for all these metals from PTMD.
 Also, for Aluminum and Copper, we have obtained isobars from NPT ensemble simulations.

\section {Simulation of Thermodynamic Properties}
\subsection{Liquid-Vapor Phase Equilibria}
We used the PTMD method developed by Lu and Hentschke\cite{Lu1,Lu2,Lu3}  to obtain the liquid-Vapor phase coexistence 
points.The basic idea of PTMD is to simulate the conditions of liquid-vapor coexistence.
The system contains two simulation boxes. Total number of particles in the system is kept constant. However, exchange of particles between 
boxes is allowed. One box is assumed to be situated in an infinite medium of homogeneous liquid at a (given) constant temperature and the 
other is assumed to be situated in an infinite medium of homogeneous vapor at the same temperature. Since the idea is to get the 
thermodynamic properties of a macroscopic system, the interface effects are neglected. The coexistence conditions are simulated by evolving the 
boxes in such a way that they have same temperature, pressure and chemical potential after equilibration as required by the Gibbs phase rule. 
Initially, each box is given a guess density(i.e., no. of particles and volume). Equilibration would be faster if the guess densities are 
closer to the coexisting liquid and vapor densities.
 Periodic boundary conditions are applied to each box to ensure that they represent the bulk coexisting phases.

 Temperature fluctuations in each box are controlled by a Berendsen thermostat \cite{Berendsen} so that they reach
 the given temperature. Pressures in both the boxes are equalized by controlling their volume fluctuations using a Berendsen 
barostat. This is done by adjusting the volume of each box such that the instantaneous pressure in one box becomes equal to 
the instantaneous pressure in the other. However the total volume of the two boxes is not restricted to be constant. The particle transfer step 
to equilibrate the chemical potentials in both the boxes is carried out after each five hundred time steps by comparing their chemical 
potentials. 
It is done as follows: A particle is chosen randomly from the box where chemical potential is more and is removed from it. Correspondingly a 
particle is introduced into the other with its potential energy calculated and the velocity taken from the Boltzmann distribution of 
corresponding temperature. Care is taken so that the introduced particle 
is not too close to any other particle in the box. With the above three 
procedures being done during the simulation, the two boxes evolve in time in such away that they have same temperature, pressure and the 
chemical potential after equilibration. Thus the system may phase separate into liquid in one box and gas in the other with proper choices of 
initial densities if the temperature of the system is less than the critical temperature. 
 Time step used in the simulations is $1$ femto second and each simulation run has
 typically $5\times 10^5$ equilibration and production 
 steps. The chemical potential has been evaluated using Widom's test particle insertion method\cite{Widom}.
 In each step $200$ test particles are inserted and the chemical potential is calculated after each five hundred steps. Total 
number of particles used in the simulation are $1728$.  
The initial densities have been chosen so that after equilibration, both the boxes contain
 a good number of particles (few hundreds) so that the averages are reliable and deviations are small.
 The method described above has been tested for the Lennard-Jones fluid initially. 
The phase diagram we obtained matched with that of the earlier simulations\cite{Pan} validating the code we developed. 
Then the code is used to obtain the LVPDs of the metals described above.

\begin{figure}
 {\includegraphics[scale=0.6,angle=-90]{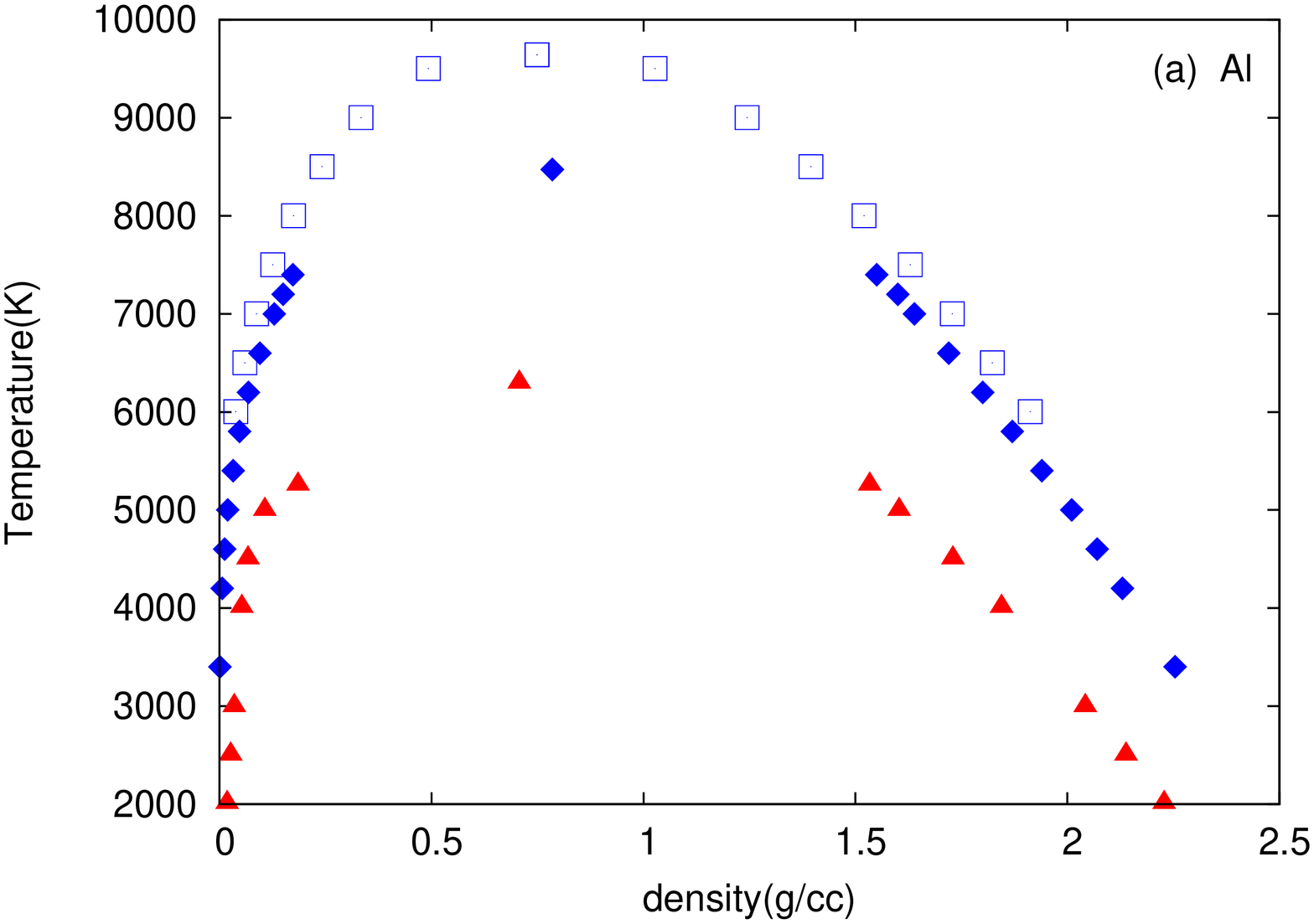}}
 \caption{\label{6} }
\end{figure}

\begin{figure}
 {\includegraphics[scale=0.6,angle=-90]{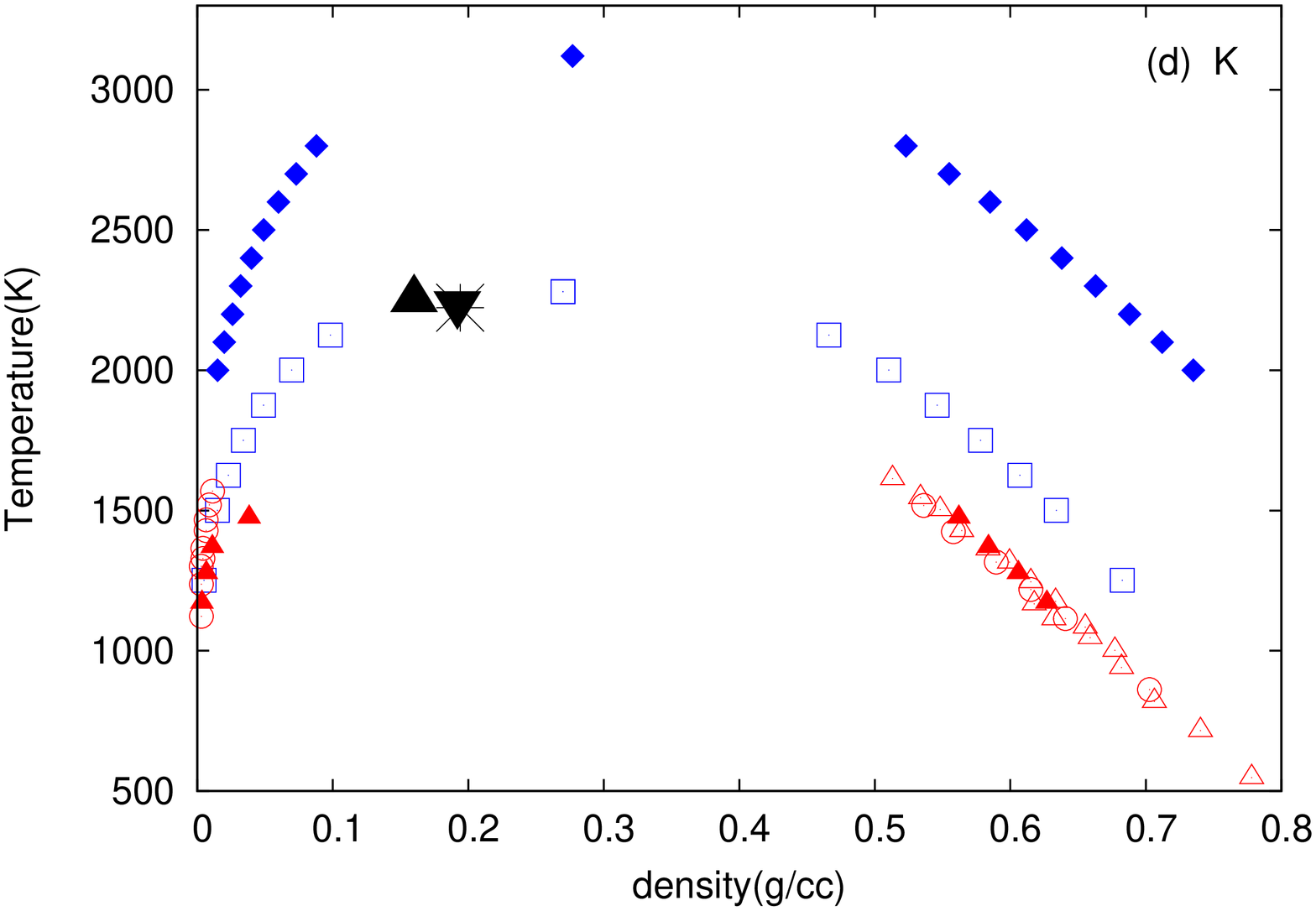}}
 \caption{\label{9}}
\end{figure}

The Temperature(T) Versus Density($\rho$) diagrams for Aluminum, Copper, Sodium and Potassium are shown in Fig.(\ref{6})- Fig.(\ref{9}) respectively.
The critical temperature($T_c$) and critical density ($\rho_c$) are obtained by fitting the simulation data to the law of 
rectilinear diameters. 
\begin{equation}
 \frac{\rho_l + \rho_v}{2} = \rho_c + A(T - T_c)
\end{equation}
and the power law 
\begin{equation}
\rho_l - \rho_v =  B(T - T_c)^\beta
\end{equation}
where $\rho_l$ and $\rho_v$ are liquid and vapor densities. $A$ and $B$ are fitting constants and $\beta = 0.33$
\cite{Alek}.

Critical parameters we obtained are compared with literature data and experiments in Table.(\ref{21}). In the
case of Aluminum, the $T_c$ and $P_c$ values we obtained are higher than the literature data.
 Whereas for copper, our data is closer to the experimental value than the available data.
 In both the cases, the LVPDs we obtained(Fig.(\ref{6}) and Fig.(\ref{7})) are closer to those obtained by Singh et. al.
\cite{Singh1}.

\begin{figure}
 {\includegraphics[scale=0.6,angle=-90]{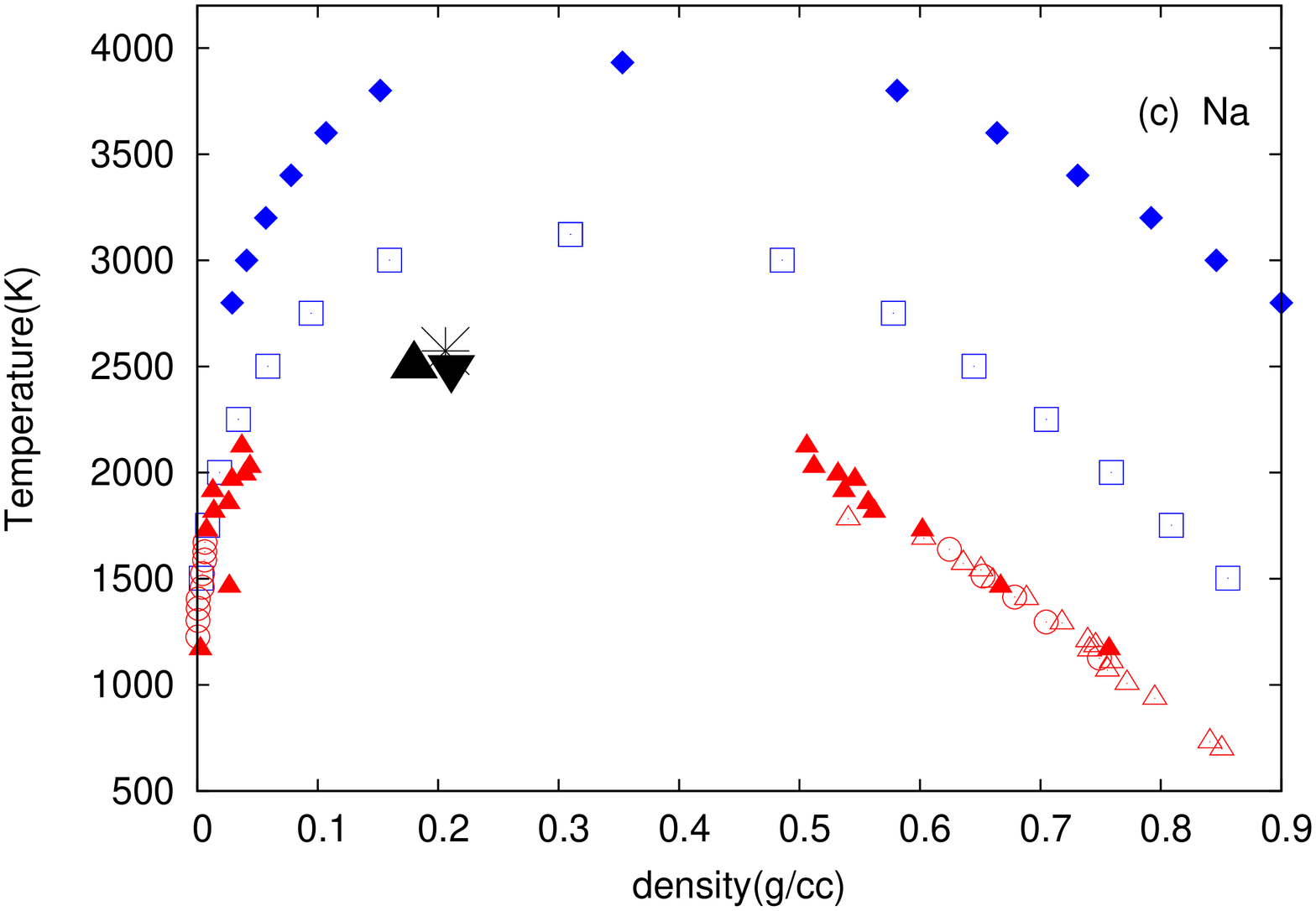}}
 \caption{\label{8} }
\end{figure}

In the cases of Sodium and Potassium,
 critical temperatures obtained using the modified potential are significantly
 improved over those obtained from Morse potential with Lincoln's 
 parameters. Also, experimental coexistence points are shown in Fig.(\ref{8}) and Fig.(\ref{9})
 for Sodium and Potassium respectively. The 
plots clearly show a significant improvement over the earlier results using Morse potential throughout the phase 
diagram. However, still there is some deviation between our results and experimental coexistence points. 

\subsection{Isobars of Aluminum and Copper}
\begin{figure}
{\includegraphics[scale=0.6,angle=-90]{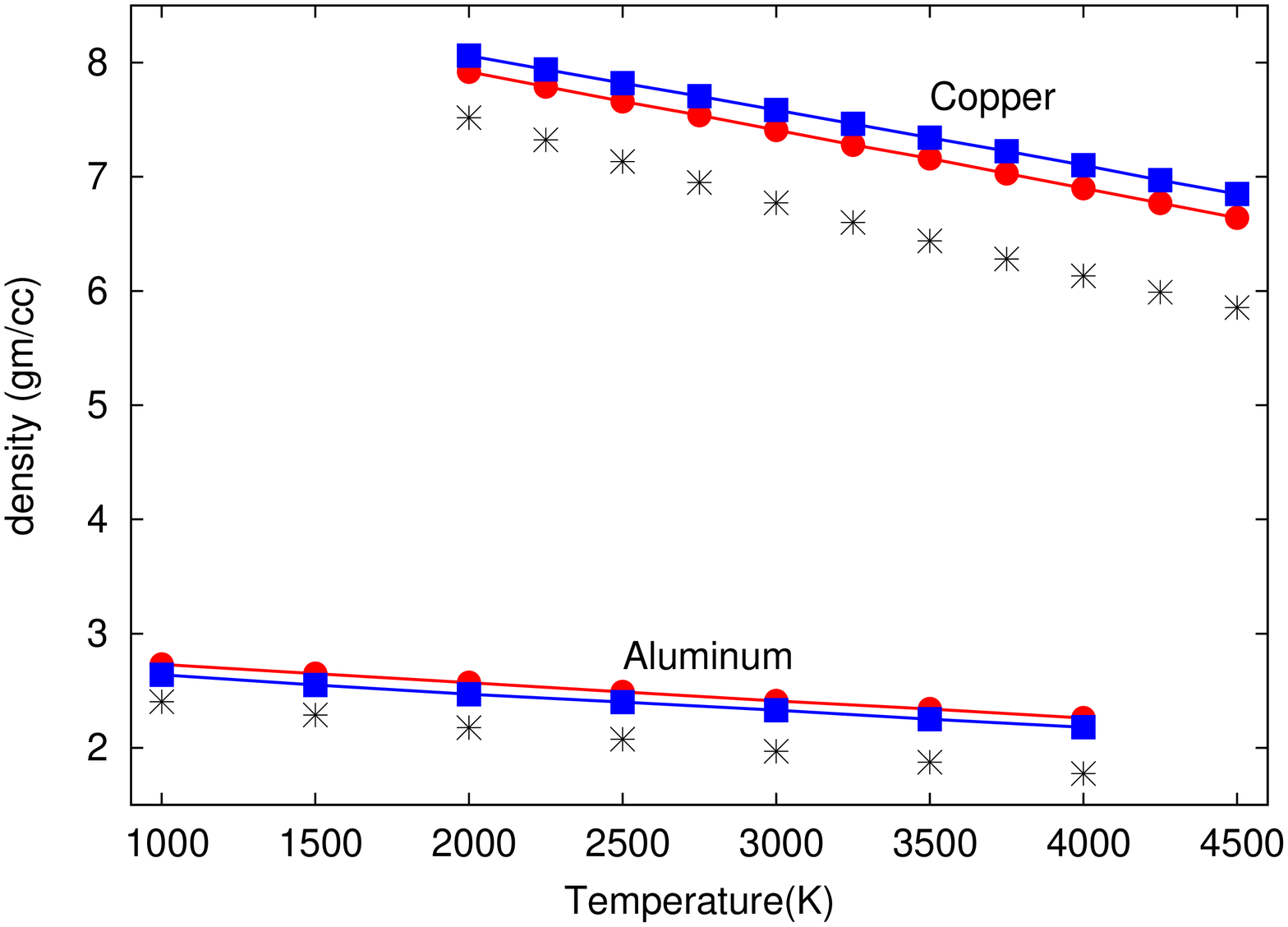}}
\caption{\label{10}}
\end{figure}

We performed NPT ensemble simulations of Aluminum and Copper using both $u_s(r)$ and $u_M(r)$ 
to obtain the isobars at 0.3GPa pressure.
Parameters obtained by Lincoln\cite{Lin} et. al. have been used in $u_M(r)$.
The simulation box contained $4000$ particles. Berendsen thermostat
and barostat have been used to control the temperature and pressure respectively.
The system is evolved in such a way that after 
equilibration it reaches to required pressure and temperature.
Typically the system is equilibrated for $20000$ time steps and then
averages are taken over $30000$ time steps. The procedure is repeated for
various temperatures keeping the pressure constant. 
The average densities at various temperatures obtained using both 
the potentials $u_s(r)$ and $u_M(r)$ are plotted in Fig.({\ref{10})}
for Aluminum and Copper. It can be seen from the figure 
that the isobars obtained using both the potentials are nearly matching for
both Aluminum and Copper. Also it can be seen that, 
in each case, the deviation of the obtained isobars from the experimental isobar is less 
at lower temperature and increases gradually as the temperature increases. In the case of Aluminum, the maximum percentage deviation is around $28\%$ and in the case of Copper, it is around $13\%$.
   
  \subsection{Results and Discussion}
\begin{figure}
 {\includegraphics[scale=0.6,angle=-90]{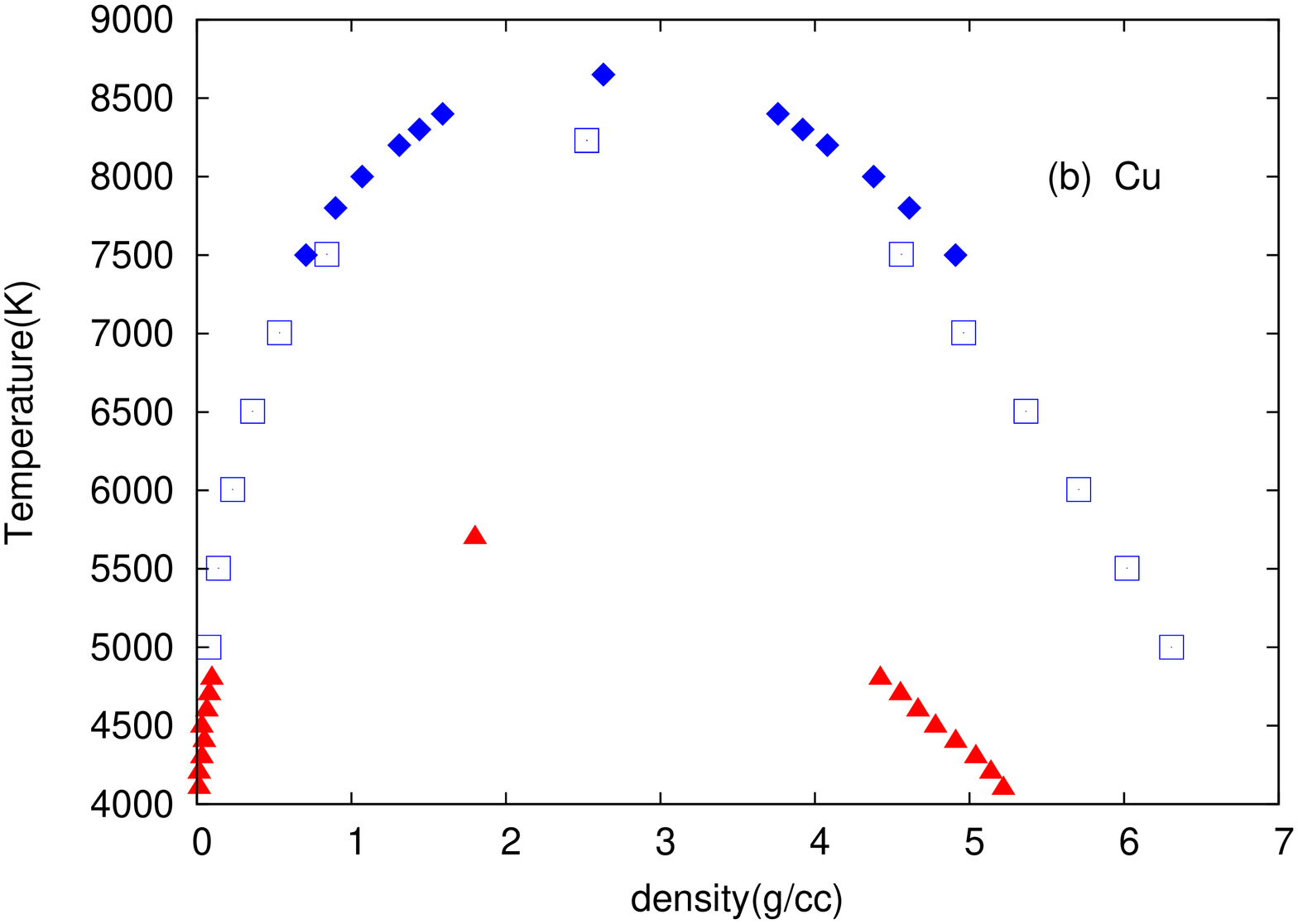}}
 \caption{\label{7} }
\end{figure}

Above results show that the critical points and LVPDs obtained using $u_s(r)$ differ from the earlier results using 
Morse potentials\cite{Singh1} in each case. 
 In the case of Aluminum, $T_c$ of present calculation is higher than that obtained by Singh et. al.
 Whereas in the case of Copper, the $T_c$ of present calculation is smaller than that obtained by 
Singh et. al. The reasons being the cohesive energy and the shape of the cold curve obtained from ab-initio 
calculations which is reflected in $U_s(V)$ and thus in $u_s(r)$. 
 In the cases of Aluminum and Copper, experimental data for LVPD is not available because of drastic temperature and
  pressure conditions involved. So we compare the present results with other simulation results available in the 
literature.
In Fig.(\ref{6}), we also show the LVPD of Aluminum obtained by Bhatt et. al.\cite{Bhatt} using the EAM potential 
through GEMC simulations.
In Fig.(\ref{7}), the LVPD of Copper obtained by Alexandrov et. al. \cite{Alek} using
 the quantum Sutton-Chen EAM potential is depicted.
It can be seen from Fig.(\ref{6}) and Fig.(\ref{7}) that there is significant deviation between the present results and 
those obtained 
 using EAM potentials. In fact, the critical points and LVPDs of the modified 
potential are closer to those of Morse potential. 
However, as shown in table \ref{21}, the various estimates of critical points based
 on experimental data, ab-initio simulations and other empirical methods show a huge scatter for both Aluminum and 
Copper. Literature survey shows that estimates of critical point of Aluminum vary from around $5500K$ to
 $9600K$\cite{Morel} and in the case of Copper, as shown in Table.(\ref{21}), the estimates vary from 
 $5100K$ and $8900K$. Lack of reliable experimental results and a huge scatter in the theoretical predictions
 does not allow a conclusion to be made about the accuracy of present results in the cases of Aluminum and Copper. 
 
 In the cases of Sodium and Potassium, even close to equilibrium
 there is a significant mismatch between $U_{ML}(V)$ and ab-initio data (and hence $U_s(V)$).
 This is because of the difference in 
 cohesive energies obtained from present ab-initio calculations 
and earlier values used by Lincoln et. al. \cite{Lin}. However, the significant
 improvement in the LVPDs over those of Singh et. al. reaffirms the accuracy of the ab-initio results.

The close agreement of isobars of Aluminum and Copper obtained using $u_s(r)$ and $u_M(r)$ shows that the results of 
$u_s(r)$ does not improve much over those of Morse potential for these metals.
  However there is a slight improvement in the case of Copper (Fig.(\ref{10})).
  
 Thus the above analysis shows that, apart from equilibrium data like cohesive energy, bulk modulus etc., the  
 accuracy of the cold curve of a particular potential away from equilibrium plays 
 a major role in determining the accuracy of the thermal properties of fluids obtained from that potential.
 The deviation of present results from the experimental values could be 
 because of various reasons. Firstly, the accuracy of potentials we obtained
 are restricted by the accuracy of the ab-initio calculations.
 Secondly, the excitation of electrons to higher levels with temperature would affect the effective potentials. This
 has not been taken into account. 
Apart from these, the method we described has been used to obtain parameters of pair potentials for metals keeping in 
view their utility in theoretical models. However, it can be used with more sophisticated potentials forms including 
many body effects like EAM potentials also. This may bring down the deviation from experimental results.

\section{Conclusion}
We described a simple way of obtaining parameters of inter-particle potentials using ab-initio cold curve.
 We also proposed a modified form of empirical pair potential for metals.
 We obtained LVPDs of Aluminum, Copper, Sodium and Potassium using the potential through PTMD simulations.
In the case of alkali metals Sodium and Potassium, a significant improvement in the phase diagrams
 over those obtained from Morse potential parametrized using equilibrium data has been observed.
 In the cases of Aluminum and Copper, we got a new set of LVPDs and critical points which are closer to those obtained 
using the Morse potentials.
 However, due to lack of accurate experimental data, a 
  conclusion cannot be reached on the accuracy of the present results in these cases. 
We also performed NPT ensemble simulations to
  obtain isobars of Aluminum and Copper. Results show that the isobar of Copper is in reasonable agreement with 
experimental data while that of Aluminum showed some deviation.
 The applications lead to the conclusion that a potential would predict accurate properties of fluids if 
  the zero Kelvin isotherm obtained from it is accurate. 
 
\section{Acknowledgements}
I thank Dr N. K. Gupta for his encouragement. Also I thank Dr. Raghawendra Kumar for his help in computation.

\newpage
\vspace{5 mm}

\noindent
$\bf{Figure\mbox{ } Captions}$

\noindent
${Figure\mbox{ } 1}$: Zero Kelvin isotherm of Aluminum obtained using various potentials. Dotted Line: cold curve of 
GLJ potential fitted to ab-initio data, double dots: cold curve of Morse potential with parameters of Lincoln et. al.\cite{Lin},
dashes: cold curve of Morse potential fitted to ab-initio data, solid line: cold curve of modified potential fitted to ab-initio data,
circles: ab-initio data. Panel (a) is for expansion phase and panel (b) is for compression phase.
\vspace{5 mm}

\begin{figure}
  {\includegraphics[scale=0.6,angle=-90]{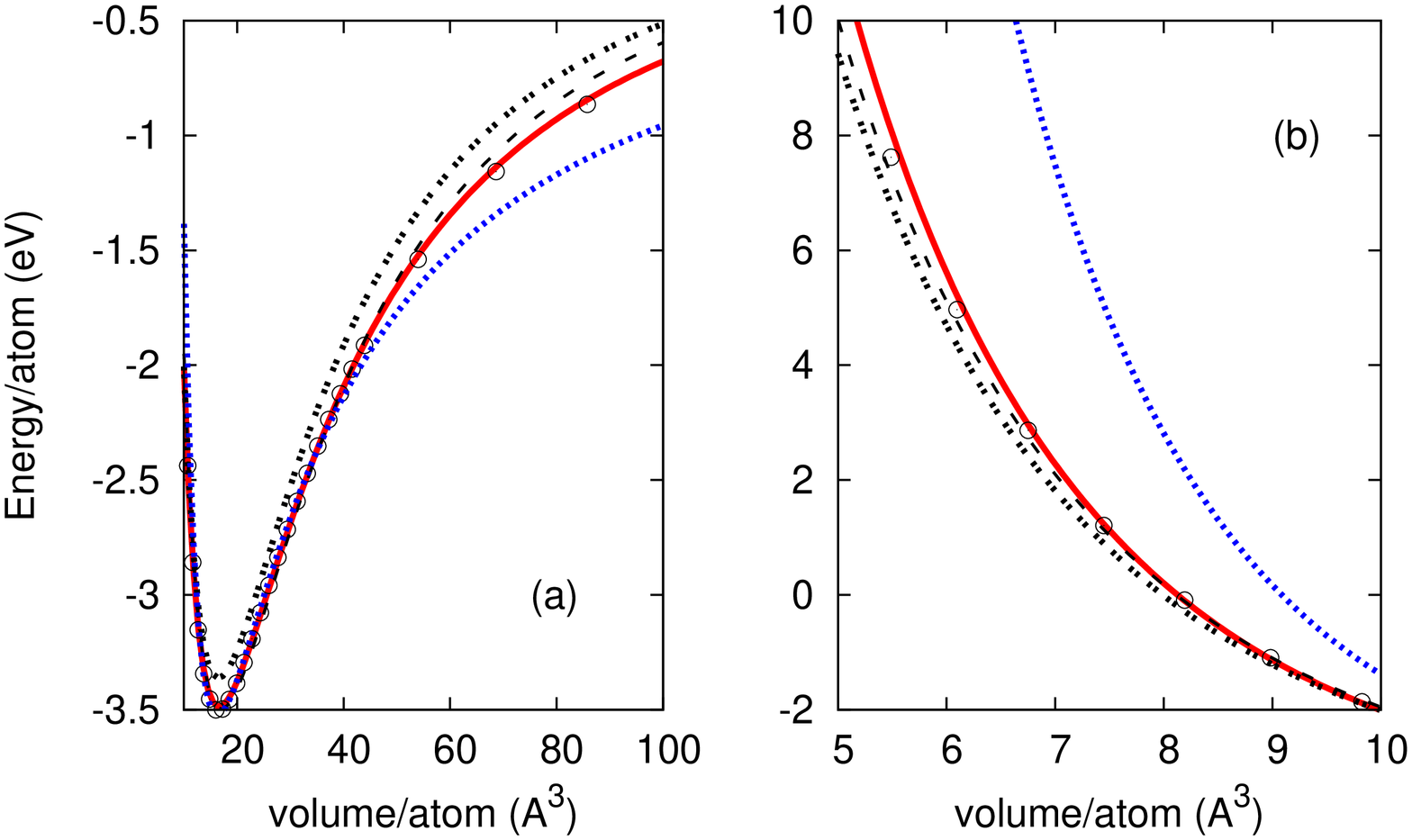}}
 \caption{\label{1}}
\end{figure}

\noindent 
${Figure\mbox{ } 2}$: Zero Kelvin isotherm of Copper obtained using various potentials. Depiction same as in Fig.(\ref{1}).
\vspace{5 mm}

 \noindent
${Figure\mbox{ } 3}$: Zero Kelvin isotherm of Sodium obtained using various potentials. Depiction same as in Fig.(\ref{1}).
\vspace{5 mm}

\noindent
${Figure\mbox{ } 4}$: Zero Kelvin isotherm of Potassium obtained using various potentials. Depiction same as in Fig.(\ref{1}).
\vspace{5 mm}

\noindent
$Figure\mbox{ } 5$:  Modified pair potentials of Aluminum, Copper, Sodium and Potassium.
\vspace{5 mm}

\noindent
$Figure\mbox{ } 6$:  Liquid Vapor Coexistence Curves of Aluminum. Squares: present work, diamonds: J.K. Singh et. al. \cite{Singh1}
, up triangles: Divesh Bhatt et. al. \cite{Bhatt}.
\vspace{5 mm}

\noindent
$Figure\mbox{ } 7$: Liquid Vapor Coexistence Curves of Copper. Squares: present work, diamonds: J.K. Singh et. al. \cite{Singh1},
up triangles: Aleksandrov et. al.\cite{Alek}.
\vspace{5 mm}

\noindent
$Figure\mbox{ } 8$:  Liquid Vapor Coexistence Curves of Sodium. Squares: present work, diamonds: J.K. Singh et. al. \cite{Singh1}, 
 filled up-triangles\cite{Dillon}, hollow circles\cite{Stone1,Stone2,Stone3} and hollow up-triangles\cite{Goltsova} are experimental data.
Experimental critical points: Up triangle\cite{Vargaftik}, down triangle\cite{Ohse}, star\cite{Dillon}.
\vspace{5 mm}

\noindent
$Figure\mbox{ } 9$:  Liquid Vapor Coexistence Curves of Potassium. Squares: Present work, diamonds: J.K. Singh et. al. \cite{Singh1}, 
 filled up-triangles\cite{Dillon}, hollow circles\cite{Stone1,Stone2,Stone3} and hollow up-triangles\cite{Goltsova} are experimental data. 
Experimental critical points: Up triangle\cite{Vargaftik}, down triangle\cite{Ohse}, star\cite{Dillon}.

\noindent
$Figure\mbox{} 10$: Isobars of Aluminum and Copper at $0.3GPa$ pressure. Lines connected with circles: obtained using modified potential,
Lines connected with squares are obtained using Morse potential with parameters of Lincoln el. al., Stars: Experimental data \cite{Gathers}. 

\newpage

\newpage
\begin{table}
\caption{\label{11} Parameters for $u_s(r)$ }
\begin{tabular}{c c c c c} \hline
metal & $\epsilon(eV)$ & $r_0(A^o)$ & $\alpha$ & $\beta$ \\ \hline
  Al& 0.220&3.568&2.499&0.7808 \\
  Cu& 0.321&2.881&3.095&0.792 \\
  Na & 0.086 & 4.567 & 3.968 & -0.5573 \\
  K & 0.088&5.425&5.172& -1.439 \\ \hline
\end{tabular}
\end{table}
\begin{table}
\caption{\label{21} Critical Point Data } 
\begin{tabular}{c c c c c} \hline
 & & & &\\
metal & $T_c (K)$ & $\rho_c (g/cc)$ & $P_c (GPa)$ & Reference \\ \hline\hline
 & & & &\\
  Al& 9643 &0.75&0.81&This work \\
    & 8472 &0.79&0.51& Singh\cite{Singh1} \\ 
    & 6299&0.71&0.88&Bhatt\cite{Bhatt}\\
    & 7963 &0.44&0.35& Faussurier\cite{Faus} \\
    & 8860 &0.28&0.31& Likalter\cite{Lik} \\
    & 8387 &0.38&0.45& Vinayak\cite{Vinayak} \\
    & & & & \\
  Cu& 8231&2.05&0.73&This work \\
      & 8650 & 2.6 & 0.95 & Singh\cite{Singh1}\\
      & 5696& 1.8 & 0.11 & Aleksandrov\cite{Alek}\\
      & 7696 & 1.93 & 0.58 & Hess(Exp.)\cite{Hess}\\
      & 8900 & 1.04 & -- & Cahill(Exp.)\cite{Cahill} \\
      & 5140 & --   & 0.4 & Martynyuk(Exp.)\cite{Martynyuk} \\
    & & & & \\
  Na & 3121 & 0.31 & 0.109 & This work \\
      & 3932 & 0.35 & 0.129 & Singh\cite{Singh1}\\
	  & 2500 & 0.18 & 0.037 & Vargaftik(Exp.)\cite{Vargaftik}\\
      & 2497 & 0.21 & 0.025 & Ohse(Exp.)\cite{Ohse}\\
	  & 2573 & 0.21 & 0.035 & Dillon(Exp.)\cite{Dillon}\\
   & & & & \\
  K & 2280&0.27&0.037&This work \\ 
      & 3120 & 0.28 & 0.053 & Singh\cite{Singh1}\\ 
	  & 2250 & 0.16 & 0.016 & Vargaftik(Exp.)\cite{Vargaftik}\\
      & 2280 & 0.19 & 0.016 & Ohse(Exp.)\cite{Ohse}\\
	  & 2223 & 0.19 & 0.016 & Dillon(Exp.)\cite{Dillon}\\
 & & & &\\ \hline
\end{tabular}
\end{table}


\begin{thebibliography}{18}

\bibitem{qeos1}
R. M. More, K. H. Warren, D. A. Young and G. B. Zimmerman, Phys. Fluids 31 (1988) 3059-3078.

\bibitem{qeos2}
T. A. Heltemes, G.A. Moses Comp. Phys. Comm. 183 (2012) 2629-2646 and references therein.

\bibitem{qeos3}
 I. V. Lomonosov, Laser and Particle Beams 25 (2007) 567-584 and references therein.

\bibitem{qeos4}
 D. A. Liberman, Phys. Rev. B 20 (1979) 4981-4989 and references therein.


\bibitem{Hansen}
J. P. Hansen and I. R. McDonald, Theory of Simple Liquids (Academic Press, London, 2006). 

\bibitem{TPT1}
A. Parola and L. Reatto, Advances in Physics 44 (1995) 211-298 and references therein.

\bibitem{TPT2}
 J.M. Bomont, Advances in Chemical Physics 139 (2008) 1-84  and references therein.

\bibitem{TPT3}
S. Zhou, Phys. Rev. E 74 (2006) 031119.

\bibitem{TPT4}
A. S. V. Ramana and S. V. G. Menon, Phys. Rev. E 87 (2013) 022101.

\bibitem{Daw1}
M. S. Daw and M. I. Baskes, Phys. Rev. B 29, (1984)6443-6453.

\bibitem{Daw2}
M. S. Daw, S. M. Foiles and M. I. Baskes, Materials Science Reports 9, (1993) 251-310 and references therein.

\bibitem{Zhou}
X. W. Zhou, H. N. G. Wadley, J.S. Filhol, and M. N. Neurock, Phys. Rev. B 69 (2004) 035402.

\bibitem{Sutton}
A. P. Sutton and J. Chen, Philos. Mag. Lett. 61 (1990) 139.

\bibitem{Cleri}
F. Cleri and V. Rosato, Phys. Rev. B 48 (1993) 22-33. 

\bibitem{Gelb}
L. D. Gelb and S. N. Chakraborty, J. Chem. Phys. 135(2011) 224113. 

\bibitem{Alek0}
T. Aleksandrov, C. Desgranges, J. Delhommelle, Mol. Sim. 38(2012) 1265-1270.

\bibitem{Bour}
E. Bourasseau, A. Homman, O. Durand, A. Ghoufi and P. Malfreyt, Eur. Phys. J. B 86(2013) 251.


\bibitem{GLJ}
J. X. Sun, H. C. Yang,Q. Wu and L. C. Cai J. Phys. Chem. Solids 63(2002) 113-117.

\bibitem{Morse}
P.M. Morse, Phys. Rev. 34 (1929) 57-64.

\bibitem{Jiu}
Sun Jiuxun, J. Phys.: Condens. Matter 17,  (2005) L103-L111.

\bibitem{sai}
A. S. V. Ramana, arXiv preprint arXiv:1308.1188 (2013).


\bibitem{Lin}
R. C. Lincoln, K. M. Koliwad and P. B. Ghate, Phys. Rev. 157, (1967) 463-466.

\bibitem{Singh1}
J. K. Singh , J. Adhikari , S. K. Kwak, Fluid Phase Equilibria 248, (2006) 1-6.

\bibitem{Lu1}
Z.-Y. Lu and R. Hentschke, Phys. Rev. E 63, (2001) 051801.

\bibitem{Lu2}
 Z.-Y. Lu and R. Hentschke, Phys. Rev. E 65, (2002) 041807.

\bibitem{Lu3}
Z.-Y. Lu and R. Hentschke, Phys. Rev. E 67, (2003) 061807.

\bibitem{Pan0}
A. Z. Panagiotopoulos, Mol. Phys. 61, (1987) 813-826.


\bibitem{Vasp1}
G. Kresse; J. Hafner, Phys. Rev. B  47, (1993) 558-561.

\bibitem{Vasp2}
 G. Kresse, Furthmuller, J. Computat Mater Sci  6, (1996) 15-60.

\bibitem{Vasp3}
 G.  Kresse, Furthmuller, Phys. Rev. B  54, (1996) 11169-11186.

\bibitem{Vasp4}
 G. Kresse,D. Joubert, Phys Rev B  59, (1999) 1758-1775.

\bibitem{Berendsen}
H. J. C. Berendsen, J. P. M. Postma, W. F. van Gunsteren, A. DiNola, and J. R. Haak J. Chem. Phys. 81, (1984) 3684-3691.

\bibitem{Widom}
B. widom J. Chem. Phys. 39, (1963) 2808-2812.

\bibitem{Pan}
A. Z. Panagiotopoulos, N. Quirke, M. Stapleton, D. J. Tildesley Mol. Phys. 63, (1988) 527-545.


\bibitem{Bhatt}
D. Bhatt, A. W. Jasper, N. E. Schultz, J. I. Siepmann and D. G. Truhlar J. Am. Chem. Soc. 128, (2006) 4224-4225.

\bibitem{Alek}
T. Aleksandrov, C. Desgranges, J. Delhommelle, Fluid Phase Equilibria 287, (2010) 79-83.

\bibitem{Morel}
V. Morel, A. Bultel and B. G. Cheron, Int. J. Thermophysics 30, (2009) 1853-1863











\bibitem{Faus}
G. Faussurier, C. Blancard, and P. L. Silvestrelli, Phys. Rev. B 79, (2009) 134202-134209.

\bibitem{Lik}
A. A. Likalter, Physica A 311, (2002) 137-149.

\bibitem{Vinayak}
V. Mishra, S. Chaturvedi, Physica B 407, (2012) 2533-2537.






\bibitem{Hess}
H. Hess, Z. Metallkd, 89, (1998) 388-393.

\bibitem{Cahill}
J. A. Cahill, A. D. Krishenbaum, J. Phys. Chem.66, (1962) 1080-1082

\bibitem{Martynyuk}
M.M. Martynyuk, O.G. Panteleichuk, High Temperature 14 (1976) 1201.

\bibitem{Vargaftik}
N. B. Vargaftik, "Tables on thermodynamic properties of liquids and gases", John Wiley \& Sons, Inc. (1975).


\bibitem{Ohse}
R. W. Ohse, J. F. Babelot, J. Magill, M. Tenenbaum, Pure \& Appl. Chem. 57, (1985) 1407-1426.


\bibitem{Dillon}
I. G. Dillon, P. A. Nelson, B. S. Swanson, J. Chem. Phys. 44, (1966) 4229-4238.

\bibitem{Stone1}
J.P. Stone, C.T. Ewing, J.R. Spann, E.W. Steinkuller, D.D. Williams, R.R. Miller,
J. Chem. Eng. Data, 11, (1966) 309-314 

\bibitem{Stone2}
J.P. Stone, C.T. Ewing, J.R. Spann, E.W. Steinkuller, D.D. Williams, R.R. Miller,
J. Chem. Eng. Data, 11, (1966) 315-320 

\bibitem{Stone3}
J.P. Stone, C.T. Ewing, J.R. Spann, E.W. Steinkuller, D.D. Williams, R.R. Miller,
J. Chem. Eng. Data, 11, (1966) 320-322.

\bibitem{Goltsova}
E.I. Gol'tsova,  High. Temp., 4, (1966) 348—351.


\bibitem{Gathers}
G. R. Gathers, Int. J. of Thermophysics 4, (1983) 209-226

\end{thebibliography}
\end{document}